\documentstyle[sprocl]{article}

\input{epsf}
\def\bfg{\begin{figure}}
\def\efg{\end{figure}}
\def\ts{\textstyle}

\def\be{\begin{equation}}
\def\ee{\end{equation}}
\def\bea{\begin{eqnarray}}
\def\eea{\end{eqnarray}}

\begin{document}

\title{Resonant Gravitational Wave Amplification - Axion and Inflaton}

\author{BRUCE A. BASSETT}

\address{International School for Advanced Studies \\Via Beirut 2-4, 
34014\\Trieste, Italy\\E-mail: bruce@stardust.sissa.it} 

\maketitle\abstracts{ 
We demonstrate the parametric amplification of the stochastic
gravitational wave  background during inflationary reheating and during
axion/moduli oscillations. This  enhances the detectability of the 
string/inflationary gravity wave signal, leaving a finger-print on the 
spectrum which might be found  with future gravitational wave detectors.}

\section{Introduction}

The nature of the cosmological background of gravitational waves is a 
subject of great interest at present, with many possible sources 
ranging from  quantum fluctuations in a string-dominated phase 
\cite{gasp97,gasp96} or inflation \cite{krau97}, to oscillations of cosmic 
string loops \cite{allen}. 

Here we focus instead on a powerful mechanism for distortion and 
amplification of any existing gravitational wave  background - namely 
damped parametric resonance due to oscillatory phases that the universe 
may have undergone. Examples are provided  by reheating at the 
end of inflation \cite{KLS94}, an oscillatory dilaton phase, or during 
coherent axion or moduli oscillations if they form a significant portion 
of the dark matter. 

We will discuss the amplification within the gauge-invariant Bardeen
formalism. Using the gauge-invariant and covariant
electric and magnetic parts of the Weyl tensor gives similar
results \cite{bass1}. The evolution of the transverse-traceless (TT) metric 
perturbations $h_{ab}$ are naturally described by the Fourier mode 
functions  $h_{\epsilon,k}$, where $\epsilon = \{+,\times\}$ are the 
polarisation states. The $h_k$  \footnote{\noindent From now on we  
surpress the  polarisation label. We also  restrict ourselves to the case of 
flat spatial sections.} satisfy:
\begin{equation}
\ddot{h}_k + 3\frac{\dot{a}}{a} \dot{h}_k + \frac{k^2}{a^2} h_k = 0\,.
\label{eq:bard1}
\end{equation}
Here $a(t)$ is the scale factor of the universe which obeys the Friedmann 
Eq.:
\be
\left(\frac{\dot{a}}{a}\right)^2 = \frac{\kappa}{3} \mu =  
\frac{\kappa}{3} \left(\frac{1}{2}\dot{\phi}^2 + V(\phi)\right) \,,
\label{eq:fried}
\ee
where $\kappa = 8\pi G$ and $\mu$ is the relativistic energy density which 
we have specified to be in the form of a scalar field 
$\phi$, with potential $V(\phi)$. This gives us enough freedom to model 
both reheating physics and the oscillations of the axion condensate.
Using Eq. (\ref{eq:fried}), we can rewrite Eq. (\ref{eq:bard1}) as  
\cite{BL97}: 
\be
\frac{ d^2 (a^{3/2} h_k)}{dt^2}  + \left(\frac{k^2}{a^2} + \ts{3\over4}
\kappa p\right)(a^{3/2} h_k) = 0\,,
\label{eq:bard2}
\ee
where $p = \dot{\phi}^2/2 - V(\phi)$ is the pressure.

\section{Parametric amplification of gravitational waves}

Now to illustrate parametric resonance, consider the $n$-dimensional  first 
order system: 
\begin{equation}
\dot{\bf y} = P(t) {\bf y}
\label{eq:sys}
\end{equation}
where $P$ is any matrix with period $T$. Then  Eq. (\ref{eq:sys}) has $n$ 
linearly independent normal solutions of the form: 
\begin{equation}
{\bf y}_i = {\bf p}_i(t) e^{\mu_i t}
\label{eq:solnsys}
\end{equation}
where the $\mu_i$ are the characteristic/Floquet exponents of the system and 
the ${\bf p}_i$ are functions of period $T$. Then the  $n$ 
characteristic numbers defined  by $\rho_i = e^{\mu_i T}$ satisfy:
\begin{equation}
\rho_1 \rho_2 ... \rho_n = \exp\left( \int_0^T Tr P(s) ds \right)
\label{eq:prodsys}
\end{equation}
with repeated characteristic numbers counted accordingly. The trace
of $P(t)$ is thus the  crucial factor determining the existence of 
exponentially amplified modes. If Tr$P(t) > 0$ then eq. 
(\ref{eq:prodsys})  implies that: 
\begin{equation}
\Pi_{i =1}^n \rho_i > 1
\label{eq:charsys}
\end{equation}
which  implies that at least one of the $\rho_i > 1 ~ \Longrightarrow ~\mu_i
> 0$ and hence by eq. (\ref{eq:solnsys}) there is at least one unbounded,
exponentially growing solution.

The archetypal example is provided by the Mathieu equation ($n = 2$):
\be
\ddot{y} + [A - 2q \cos(2 t)]y = 0
\label{eq:math}
\ee
Let $\mu = \mbox{max}\{\mu_1,\mu_2\}$. Then  $\mu$ is non-trivially 
related to the parameters $(A,q)$ which span an instability chart consisting 
of an infinite hierarchy of resonance bands  where $\mu > 0$ \cite{KLS94}. 
Defining $\epsilon = A/(2q) -  1$, we plot  $\mu$ vs. ($\epsilon,q$) 
in Fig. (1) using the piecewise quadratic approximation 
\cite{yosh95,FKYY95}.  This shows that for $\epsilon < 0$ the resonance is 
particularly strong.

The key point we wish to make is that Eq. (\ref{eq:bard2}) takes the form 
of the Mathieu equation (\ref{eq:math}) when $V(\phi) = m_{\phi}^2 
\phi^2/2$, as  appropriate for studying chaotic inflation \cite{KLS94}. In 
this  case, $\phi$ evolves as $\Phi \sin(m_{\phi} t)$. Then the pressure is: 
\be
p = - \frac{m_{\phi}^2}{2} \Phi^2 \cos(2m_{\phi}t)
\label{eq:pres}
\ee
yielding  a Mathieu Eq. with parameters:
\begin{equation}
A(k) = \frac{k^2}{a^2 m_{\phi}^2}~,~~q =  \frac{3\kappa 
\Phi^2}{16}~,~~\epsilon =  \frac{32 k^2}{3 \kappa a^2 m_{\phi}^2 \Phi^2} - 1
\label{eq:param}
\end{equation}
showing that in this case, unlike in the case of standard reheating with 
a positive coupling constant \cite{KLS94}, $\epsilon < 0$ is possible and 
gravitational wave amplification can be significant if $\Phi \sim 
M_{pl}$. 

The effect of the expansion of the universe decreases $\Phi$  
and redshifts $k$, causing a decrease of both $A$ and $q$, 
though $\epsilon$ remains roughly constant. The decrease of $q$ to below 
unity is particularly important in stopping the resonance, and there is 
thus a competition between the damping  effect of the expansion, and the 
amplification due to resonance. 

One final important point regards the validity of temporal averaging. 
With an oscillating scalar field it is common to replace $\mu$ and $p$ 
with their time-averages over an  oscillation: $\overline{\mu}$ 
and $\overline{p}$. In the case that $V(\phi) \propto \phi^2$, the average 
equation of state is that of dust, $\overline{p} = 0$. From Eq. 
(\ref{eq:bard2}) this falsely predicts that  there is no resonant 
amplification of the stochastic gravitational wave 
background during reheating or axion oscillations. This shows that 
temporal averaging is invalid. 

\begin{figure}
\epsfxsize  2.4in
\epsffile{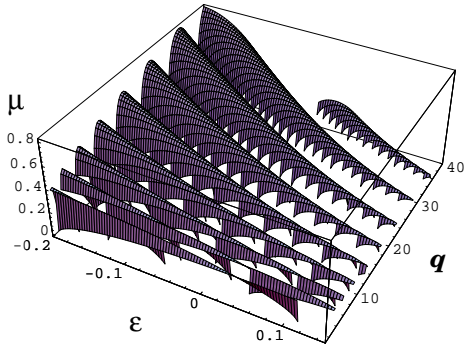}
\caption{The Floquet index $\mu$ on
the  stability-instability chart for  Eq. (8). 
Notice the rapid increase of $\mu$ for decreasing $\epsilon$ and 
increasing $q$.} 
\label{fig:floquet}
\end{figure}

\section{Applications}

\subsection{Inflationary reheating}

The gravitational wave spectrum produced during inflation is nearly 
scale-invariant. However, during reheating via a second order phase 
transition this scale-invariance is broken and the {\em rms} value of the 
spectrum is amplified. From the form of $q$ in  Eq. (\ref{eq:param}), this 
amplification is clearly strongly  dependent on $\Phi$, the initial 
amplitude of inflaton oscillations. This implies that the breaking of 
scale-invariance during reheating is much stronger in chaotic 
inflationary models than in new inflation, since $\Phi$ is much larger in 
the former case. More discussion can be found in \cite{bass1}.

\subsection{The axion and massive moduli}

The axion\cite{BS96} is an oscillating scalar field and a natural cold dark 
matter 
candidate. Unlike reheating, which lasts a very short time, the axion 
oscillations would last a large proportion of the universe's history,  
and hence might cause significant tensor amplification. The axion
pontential is given by \cite{KSS97,kim}: 
\be
V(\phi) = \Lambda^4\left[1 - \cos\left(\frac{\phi}{f_a}\right) \right]
\label{eq:pot}
\ee
with $f_a$ the axion decay constant and $\Lambda = f_a m_a$, where $m_a$ 
is the axion mass.  The standard QCD axion has $\Lambda = \Lambda_{QCD} 
\sim 200$ MeV, $f_a \sim 10^{12}$ GeV and gains a non-zero mass due to 
instanton effects at an energy around $\Lambda_{QCD}$ \cite{kim}. There also 
exist  massive moduli in  supergravity and superstring theories with much 
less  constrained  parameters, for example one may take $\Lambda \sim 
10^{16}$  GeV and $f_a \sim M_{pl}$ \cite{KSS97}, with the moduli 
generically gaining mass at the epoch of supersymmetry breaking. 

To understand the implications of axion oscillations,
let us approximate Eq. (\ref{eq:pot}) by the first, quadratic, term in the 
Taylor series. We can then use the results of Eq. (\ref{eq:param}) with 
the replacement $m_{\phi}^2 \rightarrow \Lambda^4/f_a^2$ so that roughly 
we have $A \simeq k^2 f_a^2/(\Lambda^4 a^2)$ and $q \propto \Phi^2$. 
For the values given above, this yields
\be 
A_{QCD} \sim 10^{27} \frac{k^2}{a^2}~,~~~ A_{moduli} \sim 10^{-26} 
\frac{k^2}{a^2}
\label{eq:param2}
\ee  
This implies  that massive moduli are more likely to lead to large 
amplifications of  the background gravity wave spectrum since $\epsilon = 
A/(2q) - 1 < 0$ for a huge range of modes, while in the case 
of the QCD axion,  only a tiny fraction of the modes, near $k = 0$, have 
negative $\epsilon$.  

On the other hand, only if the moduli or axions started with 
near-Planck expectation values, $\Phi \sim M_{pl}$, will there be 
significant amplification in either case.  

\section{Conclusions} 

We have shown that damped parametric resonance is important in 
understanding gravitational wave evolution 
during  phases where a significant component of the energy density 
of the universe oscillates, such as during a second order phase 
transition or if the dark matter lies in an oscillating scalar field. 

This parametric resonance amplifies the resident stochastic background, 
changing the frequency dependence of the spectrum  and 
enhancing the {\em rms} amplitude. This implies that the possibilities of 
detecting the  stochastic background of  gravitational waves may be 
better than previously thought. In addition there is the intruiging 
possibility of indirect detection of the axion or moduli via their 
finger-prints on the gravitational wave spectrum. 

\section*{Acknowledgments}
I thank Marco Serone for enlightening discussions.

\end{document}